\documentclass[fleqn,10pt]{wlscirep}
\usepackage[utf8]{inputenc}
\usepackage[T1]{fontenc}
\title{Assessing radiomics feature stability with simulated CT acquisitions}

\author[1,*]{Kyriakos Flouris}
\author[2]{Oscar Jimenez-del-Toro}
\author[3]{Christoph Aberle}
\author[3]{Michael Bach}
\author[4]{Roger Schaer}
\author[3]{Markus Obmann}
\author[3]{Bram Stieltjes}
\author[2,4]{Henning M\"uller}
\author[2,5]{Adrien Depeursinge}
\author[1]{Ender Konukoglu}
\affil[1]{Computer Vision Lab, ETH Zurich, Zurich, Switzerland}
\affil[2]{University of Applied Sciences Western Switzerland (HES-SO) Valais, Sierre, Switzerland}
\affil[3]{Department of Radiology, University Hospital Basel, University of Basel, Basel, Switzerland}
\affil[4]{Faculty of Medicine, University of Geneva (UNIGE), Geneva, Switzerland}
\affil[5]{Department of Nuclear Medicine and Molecular Imaging, Lausanne University Hospital, Lausanne, Switzerland}

\affil[*]{kflouris@vision.ee.ethz.ch}

\begin{abstract}
Medical imaging quantitative features had once disputable usefulness in clinical studies. Nowadays,  advancements in analysis techniques, for instance through machine learning, have enabled quantitative features to be progressively useful in diagnosis and research. Tissue characterisation is improved via the ``radiomics'' features, whose extraction can be automated.
Despite the advances, stability of quantitative features remains an important open problem. 
As features can be highly sensitive to variations of acquisition details, it is not trivial to quantify stability and efficiently select stable features. 
In this work, we develop and validate a Computed Tomography (CT) simulator environment based on the publicly available ASTRA toolbox (www.astra-toolbox.com).  We show that the variability, stability and discriminative power of the radiomics features extracted from the \emph{virtual phantom} images generated by the simulator are similar to those observed in a tandem phantom study. Additionally, we show that the variability is matched between a multi-center phantom study and simulated results. Consequently, we demonstrate that the simulator can be utilised to assess radiomics features' stability and discriminative power.  

\end{abstract}
\begin{document}

\flushbottom
\maketitle
%
%
\thispagestyle{empty}

\section{Introduction}


Computerized quantitative analysis of medical images is emerging as a promising approach in radiological practice and healthcare research \cite{imageanalysis1, radiomics1, radiomics2, radiomicsctcovid}. These methods extract measurements quantifying various aspects of the image that include basic intensity statistics as well as more complicated metrics quantifying spatial intensity heterogeneity. Extracted measurements are then used as image biomarkers in predicting relevant outcomes. In recent years, numerous researchers demonstrated the capability of this approach for diagnosis, stratification, and prognosis \cite{radiomics3, radiomics4}. Moreover, since the extraction of measurements as well as the prediction stage are all algorithmic, quantitative analysis is an efficient approach that can complement radiologists' visual interpretation and analysis.

Advanced artificial intelligence techniques \cite{mlbook1 }, such as deep learning, take the quantitative analysis approach one step further \cite{radiodeep,mrideep}. They remove the need to engineer measurements to extract from images for a given task. Instead, they optimize their parameters to extract task-optimal measurements and predict based on them. In the respective language, quantitative measurements are called ``features''. While the optimization requires large number of data samples, i.e., training samples, if such large datasets exist, deep learning algorithms can provide substantial accuracy gains \cite{deeplcancer}.

An important limitation of the quantitative analysis approach is its sensitivity to variations in scanning conditions \cite{radiomics_lim}. While the methods aim to extract measurements characterising the underlying tissue composition and microstructure, they are indeed measurements taken from the image, which is merely a representation of the tissue. Critically, image characteristics heavily rely on the acquisition details, e.g., resolution, radiation dose, noise, reconstruction algorithm. Depending on the properties of the algorithm and the measurement, the extracted quantities can be highly sensitive to variations in the image acquisition parameters \cite{variability1, variability2, ctvariabily}. This sensitivity inhibits the generalisation capabilities of such measurements. If acquisition details are not perfectly matched, two different images, even of the same tissue, will yield different measurements. A  number  of  studies  have  reported  the  impact  on  CT  radiomics  analysis  caused  by the  variability  of  acquisition  parameters  and  post-process  variables \cite{analct1,analct2, analct3, analct4}. Any algorithm or analysis based on these measurements will therefore not be reliable for use with unseen scanners. 

The ideal way to study the sensitivity of measurements is to perform test-retest studies \cite{teststudies}. This would comprise of imaging a group of subjects imaged under different acquisition details. To study sensitivity of a measurement, values extracted from corresponding images would be compared. When new measurements or new algorithms to extract measurements are proposed, they would be studied the same way. As this is not feasible for various imaging modalities, such as Computed Tomography (CT) due to the radiation exposure of patients in these studies, anthropomorphic printed phantoms have been proposed for CT variability studies \cite{printedphantom1, printedphantom2, printedphantom3}.

Phantom studies have been successfully used for various imaging modalities. Especially for CT, advances in 3D printing technologies allow printing volumetric patient images using materials with attenuation properties comparable to human tissue. Recent work reported variability studies using such phantoms \cite{oscar1, phantomandvariabiliy,phantomprinting}. 

While phantoms make it possible to study variability without imaging cohorts, they still require acquiring and imaging phantoms. This can be costly as well as resource and time consuming. In this work, we study whether sensitivity analysis using advanced in-silico CT simulators can yield similar results to real phantom imaging studies. To this end, a CT-scan simulator environment was set up using the publicly available \cite{astra1, astra2} ASTRA toolbox (www.astra-toolbox.com). Using a high-dose CT-image as input, the simulator outputs raw projections, which can be manipulated accordingly. For example, stationary and uncorrelated noise can be added. Additionally, the simulator allows for some freedom in geometrical parameters such as the number of projections, slice thickness, and distances. The CT-image can be reconstructed with a variety of algorithm choices, e.g. filtered back-projection and simultaneous iterative reconstruction technique.

The method is compared with an empirical anthropomorphic phantom variability study published in \cite{oscar1}. In this unique setup, the simulated phantom study is performed using the same original image from which the anthropomorphic phantom was printed and the study in~\cite{oscar1} conducted. In a sense, this can be viewed as the theoretical replication. The simulator environment was implemented to reconstruct images at different noise levels, reconstruction algorithms, and number of projections. To mimic repetition and introduce variability, each simulation parameter set was repeated via a variation of the Poisson noise random seed. For the simulated images, radiomics features were extracted and analysed. As the same source image is used for both the empirical phantom study and this work, direct comparison of the results of sensitivity analyses is possible.

The next section describes the CT simulator environment method including a brief introduction of the anthropomorphic phantom and the phantom study. In the results section a comprehensive validation and comparison of the simulator with respect to the phantom study is presented. Furthermore, a stability and discriminative power analysis and discussions can be found in the same section. The paper is summarised in the conclusions section.

\section{Method}

First, we introduce the details of the novel anthropomorphic phantom created for the tandem phantom study~\cite{oscar1}. A high dose CT-scan of this phantom is used as the simulator input. Second, the extracted radiomics features, the principal component analysis and the simulator environment are described in detail.

\subsection{Anthropomorphic phantom and phantom CT acquisitions}
Here, we provide brief details of the anthropomorphic phantom study presented in~\cite{oscar1} for completeness. For further details, we refer the reader to the original publication. 

A realistic radio-opaque three-dimensional phantom was designed from real patient CT data. Namely, the compilation of a half-mirrored lung including a tumor and an abdominal liver section with a metastasis from a colon carcinoma~\cite{oscar1}. The phantom was manufactured via stacking sheets of printed aqueous potassium iodide solution on paper \cite{phantom1}. The lung tumor section is a replication of a publicly available patient data set for radiomics phantoms, from the Image Biomarker Standardization Initiative \cite{biometrics}. The lung section was neither used in this work nor the tandem phantom CT study. Tissue equivalent attenuation at a defined energy spectrum was calibrated at 120 kVp. The contrast resolution of the printing technique in the phantom goes from -100 to 1000 Hounsfield units (HU). Overall, no structures can be represented whose HU is below this paper-induced threshold.  To test the contrast resolution, a circular intensity ramp was printed in the phantom running through an HU range of 0 to 1000. A reliable resolution of 2 HU difference was achieved. Consequently the abdominal region was adequately depicted for a quantitative analysis within the printed HU range.

The phantom was imaged with a Siemens SOMATOM Definition Edge  CT scanner (SSDE). To define the acquisition and image reconstruction parameters, a survey of clinical CT protocols was performed including 9 radiological institutes.
All the CT scans in that study were acquired with the same acquisition
parameters, which resulted in an approximate CT dose index of 10mGy. Namely, a tube voltage of 120 kVp, a helical pitch factor of 1.0, a 0.5 second rotation time, and a tube current time product of 147 mAs. No automatic tube current modulation was used.  

Typical reconstruction parameter settings for clinical protocols in thoracic and abdominal oncology were varied for the phantom study as follows: Reconstruction algorithm, iterative reconstruction (IR) or filtered back projection (FBP); reconstruction kernel, 2 standard soft tissue kernels per algorithm; slice thickness in millimeter, 1, 1.5, 2, 3; and slice spacing in millimeter, 0.75, 1, and 2. Series reconstructed with an IR algorithm used an ADMIRE (advanced modeled iterative reconstruction) at strength level 3. In total, 8 groups of parameter variations were selected for the phantom study to assess their impact on classic radiomics features. Initially, 20 repetition scans were performed without re-positioning of the phantom, followed by 10 repetitions with re-positioning between each measurement.
Therefore, 30 distinct acquisitions were performed for each of 8 parameter variation groups.

\begin{figure}[ht]
\centering
\includegraphics[width=0.6\linewidth]{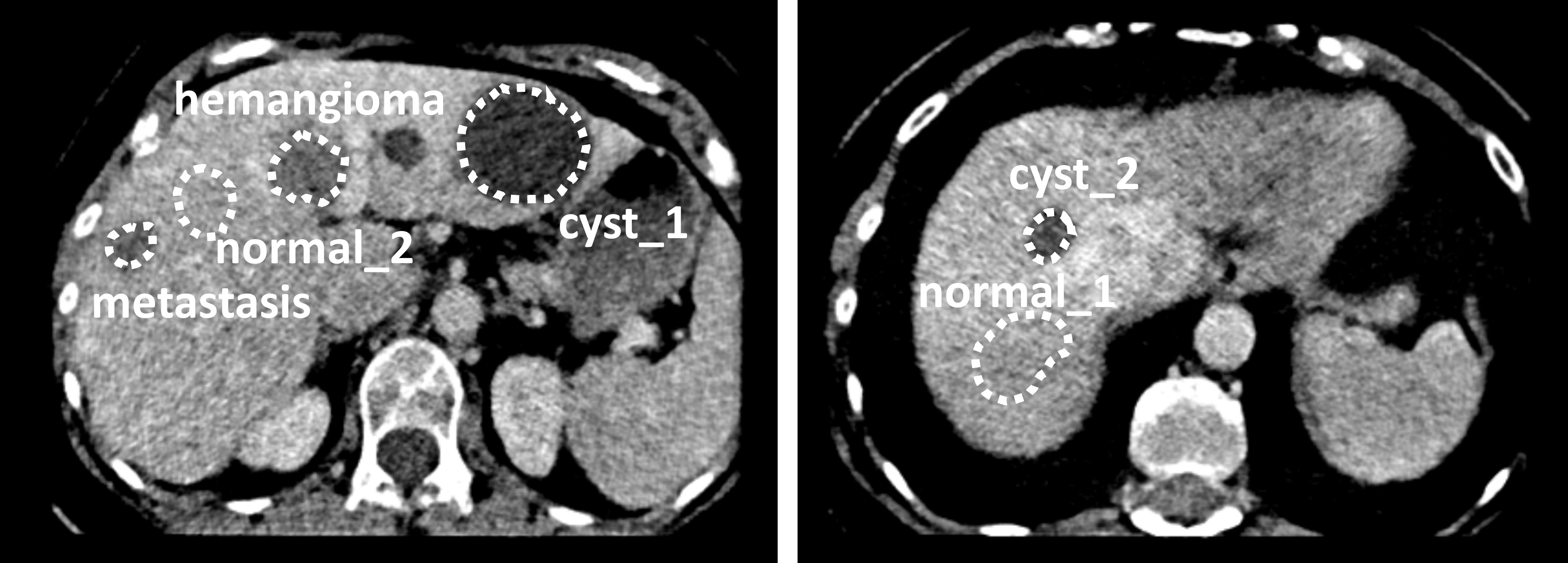}
\caption{Annotated regions of interest on the anthropomorphic phantom.}
\label{fig:rois}
\end{figure}

In the abdominal section six 3D regions of interest (ROIs) were manually annotated by a board-certified radiologist using a thin-section phantom series with 2 mm slice thickness and 1 mm spacing. The ROIs were annotated conservatively, well within the margins, thus no cross-check step of the annotations was performed by other radiologists. A polygonal outline was used on all slices individually to define the ROIs. The six ROI binary masks were stored in a 3D volume NIfTI format. Two normal liver tissue regions, two cysts, a hemangioma, and a liver metastasis from a colon carcinoma were included during the annotation process, regions can be found in Figure \ref{fig:rois}. Further details of the annotated regions and the 8 variation groups can be found in Jimenez-del-Toro et al. \cite{oscar1}.

 A multi-center phantom CT study was also carried out with 13 different scanners at selected locations in Switzerland. The scanners used were two Siemens SOMATOM Definition Edge,  two	Siemens SOMATOM Definition Flash,	a Siemens SOMATOM Edge Plus, a	Siemens SOMATOM X.Cite, a	Philips Brilliance iCT, a GE BrightSpeed S, a Philips Brilliance CT 64, a	GE Revolution Evo, a 	GE Revolution Apex,	a Canon Aquilion Prime SP and a Canon Aquilion CXL. The same protocol was implemented (as closely as possible) in all acquisitions.  A tube voltage of 120 kVp, a helical pitch factor of 1.0 and a 0.5 second rotation time were used. The tube current time product was adjusted accordingly to achieve the required dose of 10mGy. The IR reconstruction algorithm was used with slice thicknesses 2 or 2.5 millimeter and slice spacing 1 or 1.25 millimeter.

\subsection{Radiomics feature extraction and principal component analysis}\label{sec:features}
From both the simulated and phantom CT scans, a total of 86 radiomics features were extracted in 3D from the manually segmented ROIs using the open source Pyradiomics python toolkit \cite{biometrics}. Definitions for the radiomics features are available in the Pyradiomics documentation online (https://pyradiomics.readthedocs.io/en/latest/features.html). The 86 features extracted include 18 first-order statistics, 22 grey level co-occurrence matrices, 14 grey level dependence matrices, 16 grey level run length matrices and 16 grey level size zone matrices, as described briefly in the Appendix \ref{app:radiomicsdef}. Radiomics features parameters were set to their default values. More specifically, no filter was applied to the input image and a fixed bin width of 25 was used for the discretisation of the image grey level. Fixed bin size discretisation is defined such that a new bin is assigned for every intensity
interval within the bin width starting at the lowest occurring intensity.
Additionally, no normalization,  no spatial resampling, no resegmentation were performed and no HU cutoffs were used within the ROIs for the extraction. The distance between the center voxel and the neighbor, for which angles should be generated, was set to one pixel. Furthermore, for the first order radiomics the voxel array shift parameter was set to zero, for the grey level co-occurrence matrices the co-occurrences was assessed in two directions per angle, which results in a symmetrical matrix and for the grey level dependence matrices no cutoff value for dependence was set, i.e.  a neighbouring voxel was always considered  independent.

For the phantom CT acquisitions, an analysis was carried out via the principal component analysis (PCA). The first two principal components of the 86 radiomics features from all 240 phantom CT acquisitions are shown in Figures \ref{fig:pcasirt} and \ref{fig:pcafbp} with black markers. The ROIs can be  separated into 4 distinct tissue classes, i.e.  normal liver tissue, cyst, hemangioma, and liver metastasis. The differences between the four ROI classes (inter-class variation) are larger than all CT parameter variations (intra-class variation). ROIs from the normal liver tissue class are closer in the feature space than those from the other classes. All four classes remain linearly separable despite the CT parameter variations. 


Furthermore, the Wilcoxon statistic $W$ was used to assess the stability and discriminative power of isolated radiomics features \cite{oscar1}. We set a threshold of $W$ < 1 to indicate a stable comparison. The top 10 ranked features  of the phantom CT acquisitions are shown on the right-hand of the appendix Table \ref{tab:topfeatures}.

\subsection{CT simulator}

The simulator environment was implemented to reconstruct images at different noise levels, with different reconstruction algorithms, and number of projections. Each simulation parameter set was repeated ten times for different noise random seeds to approximate repeated scans. For the simulated images, feature values were extracted and analysed. Specific features are explained in detail in Section~\ref{sec:features}.

The ASTRA toolbox CT-scan and reconstruction simulator \cite{astra1} was employed for the purpose of this  study. The simulator is based on simple geometric principles for the creation of projection data (sinograms). These sinograms can then be manipulated to mimic more realistic scenarios, for example through adding Poisson noise. Subsequently, the processed images are passed to the reconstruction algorithm. To match the simulator to the phantom acquisitions, a helical scanning sequence of pitch one was realised by explicitly specifying a sequence of helical projection vectors. These explicit projection vectors define the scanning frequency, i.e. the total number of projections.
A conical beam is utilized and the target and detector are placed at 500mm and 1000mm respectively to approximate the real scanner geometry. A flat square detector of 512 by 512 of continuous pixels (1mm) was implemented for simplicity. The number of detector pixels is higher than for a clinical CT scanner (approximately 1000 by 64) but is nevertheless compensated by an equivalent decrease in the scanning frequency, making the simulations simultaneously efficient and realistic. 


Random uncorrelated noise is added at the projection level by sampling from a Poisson distribution,
\begin{equation*}
    f(k; \lambda)=\frac{\lambda^k e^{-\lambda}}{k!},
\end{equation*}
where, $f(k; \lambda)$ describes the probability of k occurrences and $\lambda$ is both the expectation and the variance of the distribution. A background intensity $I_0$ is used to define the noise level  i.e. at each pixel of the projection images:
\begin{align*}
    I_{sampled} \sim f(\lambda=I_0 e^{-I_{image}}), \\
    I_{final}= -\log(I_{sampled}/I_0).
\end{align*}
$I_{image}, I_{sampled},  I_{final}$ represent the initial image, sampled and final intensities respectively. Therefore the background intensity is inversely related to the Poisson noise. Here we denote the added noise level $I_0^{-1}$ as $A$.  The noise is added using the ``add\_noise\_to\_sino'' function in the ASTRA toolbox. 
    
\begin{figure}[ht]
\centering
\includegraphics[width=0.6\linewidth]{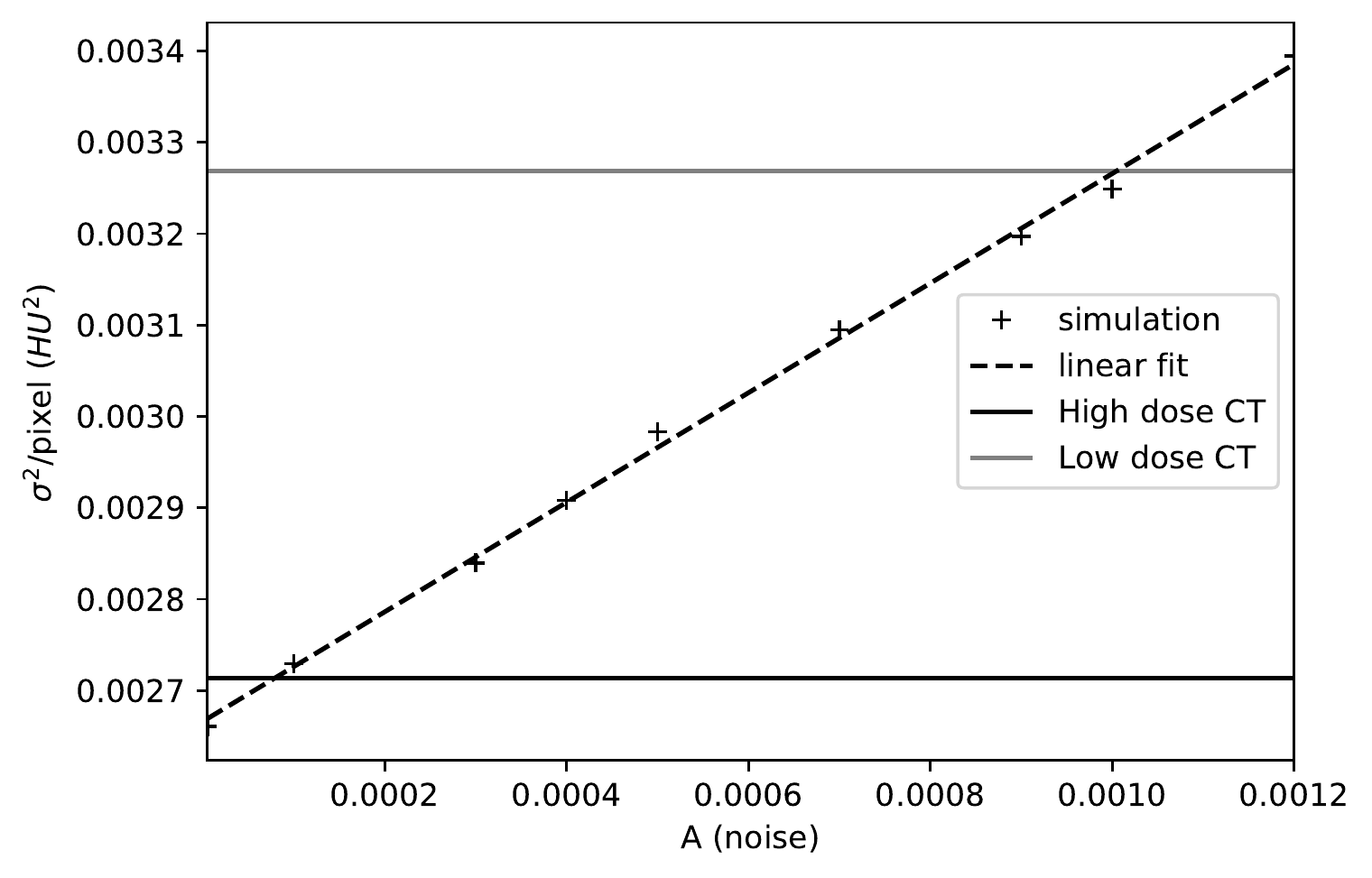}
\caption{Average pixel-wise variance of the iterative method simulated image plotted against the arbitrary noise measure. Black and grey lines denote the average variance of the high dose and low dose acquisitions.}
\label{fig:noise}
\end{figure}

To calibrate an appropriate noise level $A$ the average pixel-wise variance $\sigma^2$ is calculated for a range of $A$s and compared to the $\sigma^2$ of the phantom CT acquisitions, see Figure \ref{fig:noise}. The average $\sigma^2$ of the low dose (1 mGy) and high dose  (10 mGy) acquisitions are plotted as the horizontal lines. An approximate linear relation is observed between $\sigma^2$  and $A$ as seen from the linear fit. The crossing points between the horizontal line limits and the fitted line serve as a guide for a realistic $A$ parameter range. In the simulation study, noise levels close to the 10 mGy were used as this was the dose level used in the tandem phantom study.

The reconstructions are performed with the simultaneous iterative "SIRT" and filtered back-projection "FBP" 3D algorithms as implemented in the ASTRA toolbox. Specifically the "$SIRT3DCUDA$" with 500 iterations and "$FDKCUDA$" were used, the reconstruction kernels are fixed by the simulator and the slice thickness is the same as the pixel resolution, i.e. 1 mm. Furthermore, a distinct numerical random seed is used for the Poisson noise, to imitate repetitions as performed for the phantom CT acquisitions \cite{oscar1}. The method is very efficient numerically, as total computational time on a modern GPU is in the order of minutes per complete reconstruction.  

\begin{figure}[ht]
\centering
\includegraphics[width=\linewidth]{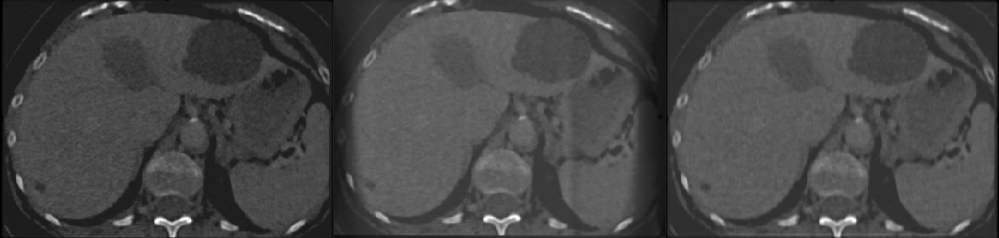}
\caption{Axial views of anthropomorphic radio-opaque phantom. Left, original input. Middle, filtered back-projection reconstruction, right, iterative reconstruction, both obtained by the CT simulator.  }
\label{fig:scans}
\end{figure}

\begin{table}[ht]
\centering
\begin{tabular}{|l|l|l|}
\hline
Parameter & range & optimal \\
\hline
Noise level ($\sigma^2/$pixel in  HU$^2$) & 2.5$\times10^{-3}$ - 2.8$\times10^{-3}$ & 2.5$\times10^{-3}$ \\
\hline
Number of Projections & 150-450 & 450 \\
\hline
\end{tabular}
\caption{\label{tab:parameters} Parameter choice for the simulation environment.}
\end{table}

\section{Results and Discussion}

Pilot simulations are carried out with the optimal set of parameters as seen in Table \ref{tab:parameters}, i.e. minimum noise and maximum number of projections for ten repetitions. First, the procedure is verified qualitatively by visual inspection of the reconstructions with optimal parameter choice, axial snapshots can be seen in Figure \ref{fig:scans}.  Both reconstruction methods are sufficiently successful. The iterative reconstruction has low noise and no artifacts are visible. The FBP method is marginally noisier and exhibits some minor artifacts, these differences are expected as theoretically the iterative method is superior, albeit more computationally expensive. 


Additionally, the Wilcoxon statistic $W$ is employed to analyze the stability and discriminative power of the radiomics features as extracted from the simulated CTs. To this end, a study is carried out to mimic the phantom CT acquisitions. Namely the simulations are separated into 8 distinct groups with different projection number and reconstruction algorithms, see Table \ref{tab:simgroups}. Within each group, repetitions are achieved via a different Poisson noise random seed. Across the study, the same noise level was added at the projection stage. The ROIs are separated into 4 distinct tissue classes, i.e. normal liver tissue, cyst, hemangioma, and liver metastasis. The analysis aims to quantify stability and discriminative power of features across parameter groups using the class definitions.

\begin{figure}[h]
\centering
\includegraphics[width=0.7\linewidth]{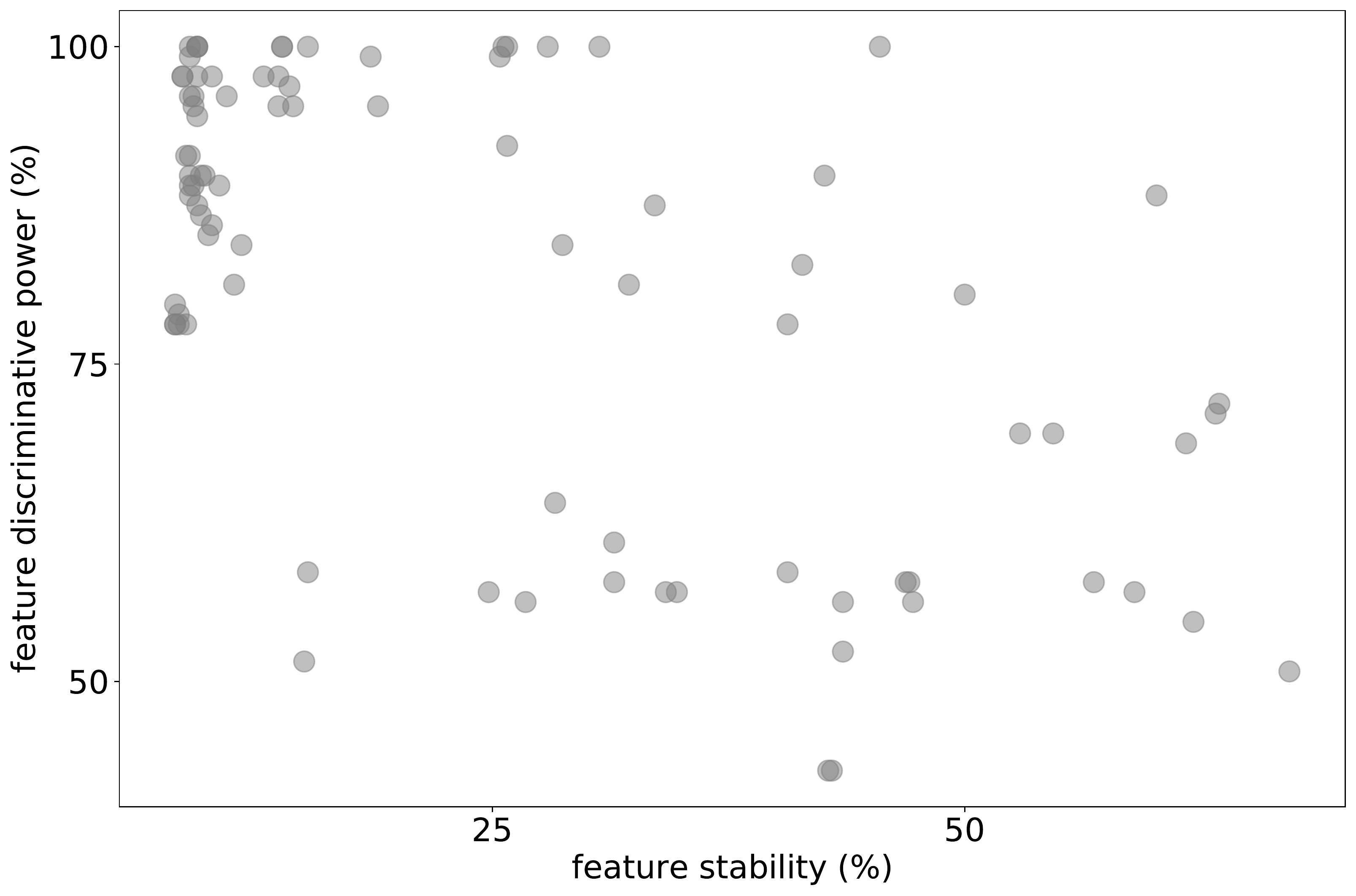}
\caption{Percentage stability of features as intra-class comparison and discriminative power inter-class comparison.}
\label{fig:stability}
\end{figure}

The result is depicted in Figure \ref{fig:stability}. The stability (intra-class variation) percentage is calculated from a pairwise comparison among the 8 parameter variation groups. This process is repeated for all available tissue classes, while all other CT parameters are kept constant. Expressly, for each feature from each class,
$W$ is calculated in-between the groups. To this end, a threshold of correlation is predetermined for $W$ at $1$. I.e.  the repetitions within the two tested groups in question follow the same distribution if $W < 1 $ and the pairwise comparison is considered successful. The percentage is calculated as the total fraction of the successful pairwise comparisons for each feature. The discriminative power (inter-class variation) is calculated via pairwise comparison in-between tissue classes for each feature and group. In contrast to stability, here a successful comparison is achieved if $W > 1$. Again the percentage represents the fraction of successful comparisons.

\begin{table}[ht]
\centering
\begin{tabular}{|l|l|l|}
\hline
Group & Reconstruction & Projections \\ 
\hline
1 & SIRT & 150 \\ 
2 & SIRT & 200 \\ 
3 & SIRT & 250 \\ 
4 & SIRT & 300 \\ 
5 & FBP & 150 \\ 
6 & FBP & 200 \\ 
7 & FBP & 250 \\ 
8 & FBP & 300 \\ 
\hline
\end{tabular}
\caption{\label{tab:simgroups} Parameter choice for the stability and discriminative power study. For all simulations, the noise level was set to $A=0.0001$, i.e equivalent to approximately 10mGy dose, and 10 different random seeds were used to achieve repetitions within the group.}
\end{table}

\begin{figure}[h!]
\centering
\includegraphics[width=0.7\linewidth]{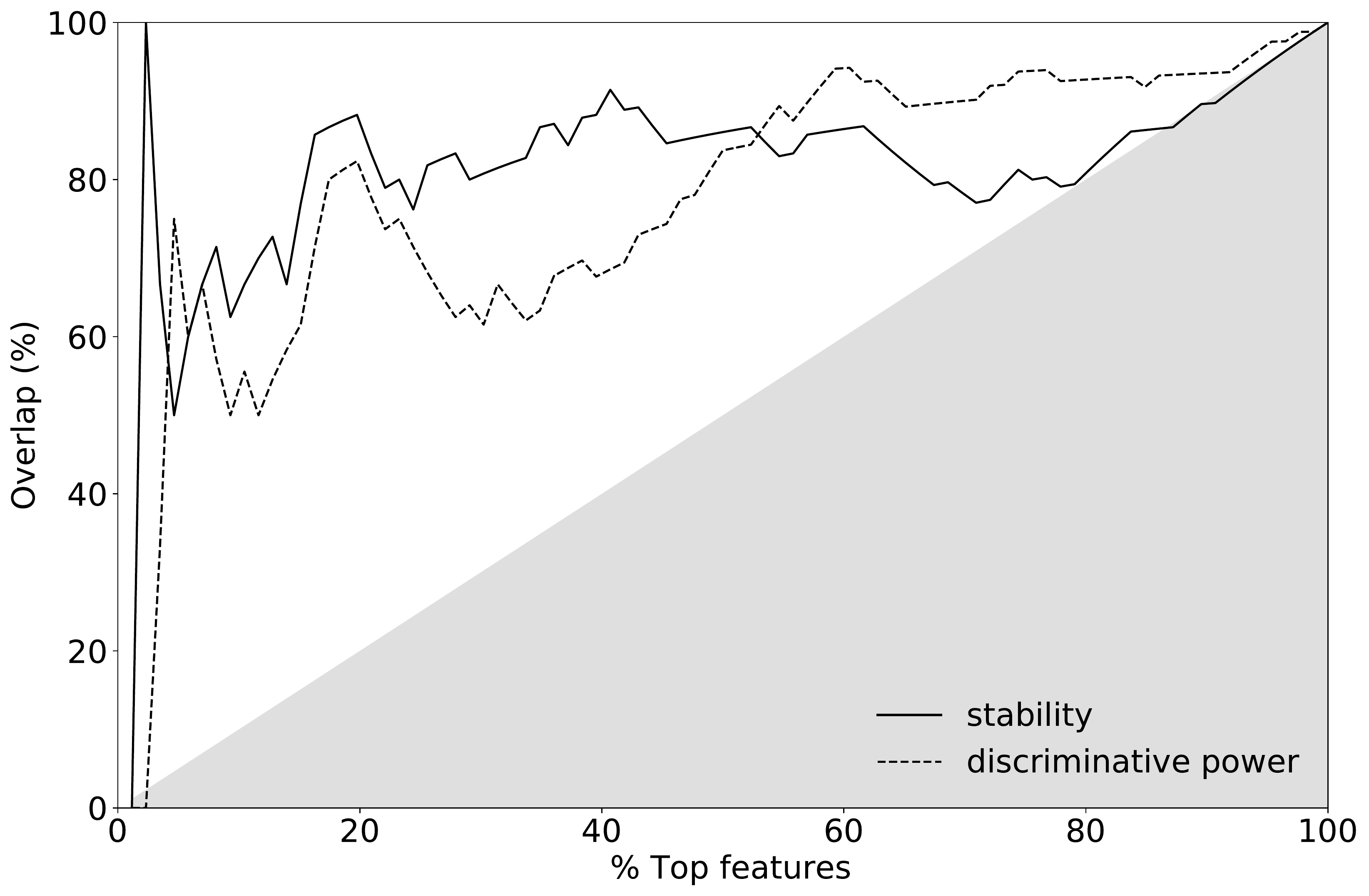}
\caption{Overlap of highest scoring features between simulation and phantom CT acquisitions plotted against ascending percentage that are considered highest scoring. Plotted for the stability and discriminative power alike. The grey area represents un- or negatively-correlated overlap between the two methods.}
\label{fig:overlap}
\end{figure}

The results show that although the majority of radiomics features had low stability for CT parameter variations, as has been previously shown in other studies \cite{radiomicsunstable1,radiomicsunstable2, radiomicsunstable3, radiomicsunstable4}, the discriminative power is high in the task of differentiating in-between the tissue classes. This relation is again observed in the phantom study that is mimicked \cite{oscar1}. The top ten features across each axis selected by the simulation environment, i.e., virtual phantom, and the phantom CT acquisitions are compared in Table \ref{tab:topfeatures}. To demonstrate the ability of the simulation environment to predict stable features, an overlap of the best scoring features relative the phantom CT acquisitions is plotted in Figure \ref{fig:overlap}. The $x$-axis represents an ascending percentage of features that are considered as the highest scoring group (e.g. $10\%=$ top 9 out of 86 features)  and the $y$-axis the percentage within that group that overlaps with the top features seen in the phantom study, http://links.lww.com/RLI/A632. For both stability and discriminate power, the overlap is consistently high, i.e., not in a linear relationship as expected for non-correlated lists.  

\begin{figure}[h!]
\centering
\includegraphics[width=0.7\linewidth]{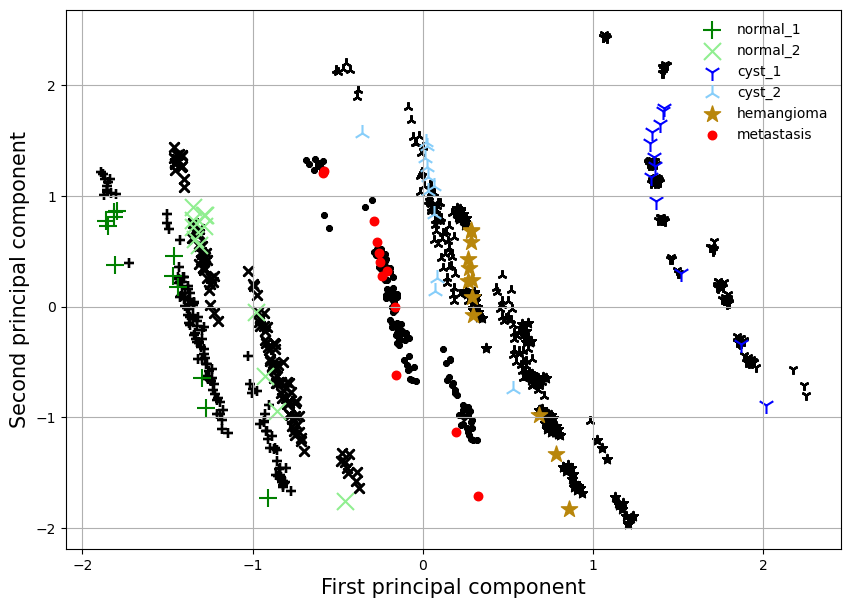}
\caption{Principal component analysis. The black markers indicate the empirical phantom study data of the region of interest with the same shape as the colored markers. The colored markers indicate the equivalent result of the simulator with iterative reconstruction  and the parameter range shown in Table \ref{tab:parameters}. The average value and variability of the two principal components are closely matched.  }
\label{fig:pcasirt}
\end{figure}

Furthermore, the radiomics features of the simulator are compared to the empirical phantom acquisitions in Figure \ref{fig:pcasirt} in a variability analysis. To this end, the principal components are calculated to investigate the similarity and  variability of the radiomics, and we use the parameter range as shown in Table \ref{tab:parameters}. As seen from Figure \ref{fig:pcasirt}, the simulation radiomics variability is in agreement with the empirical results. It should be noted here that the study was carried out in a semi-blind methodology, i.e. after matching all the possible parameters to reality, the best possible values were used to create the optimal reconstruction. Afterwards, an appropriate noise level was chosen using Figure \ref{fig:noise} for the purpose of this variability study.  

\begin{figure}[h!]
\centering
\includegraphics[width=0.7\linewidth]{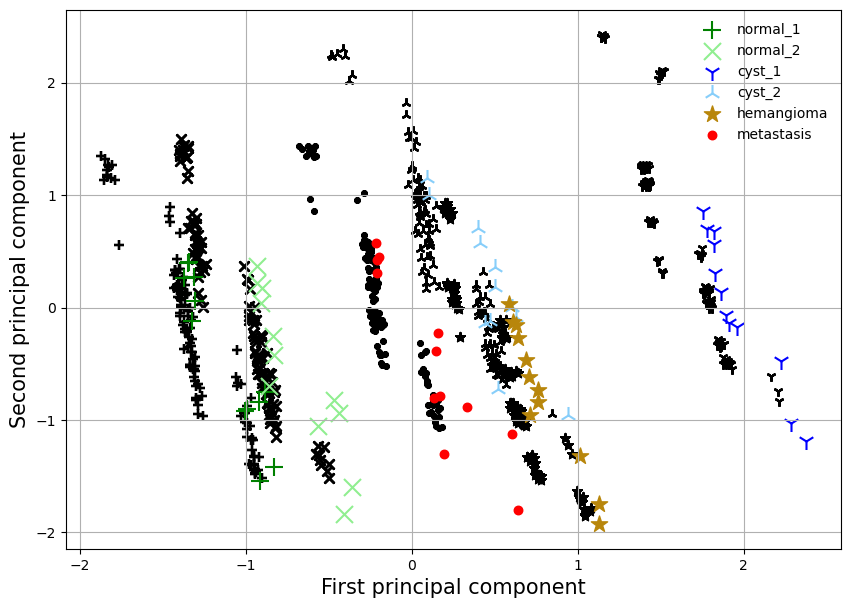}
\caption{FBP principal component analysis. The black markers indicate the empirical phantom study data of the region of interest with the same shape as the colored markers. The colored markers indicate the equivalent result of the simulator with the filtered back-projection reconstruction  and the parameter range shown in Table \ref{tab:parameters}. The average value and variability of the two principal components are matched up to a shift of the first principal component. }
\label{fig:pcafbp}
\end{figure}

The filtered back-projection method creates an inferior reconstruction as seen in Figures \ref{fig:scans} and \ref{fig:pcafbp}. There is a larger discrepancy between empirical distributions and distributions obtained through the simulation. 
Nevertheless, when the PCs are plotted the results indicate that the variance is well within the experimental result. There is a lateral shifting of  the first PC. The variability is well captured by the simulator for all six ROIs.


\begin{figure}[h!]
\centering
\includegraphics[width=0.7\linewidth]{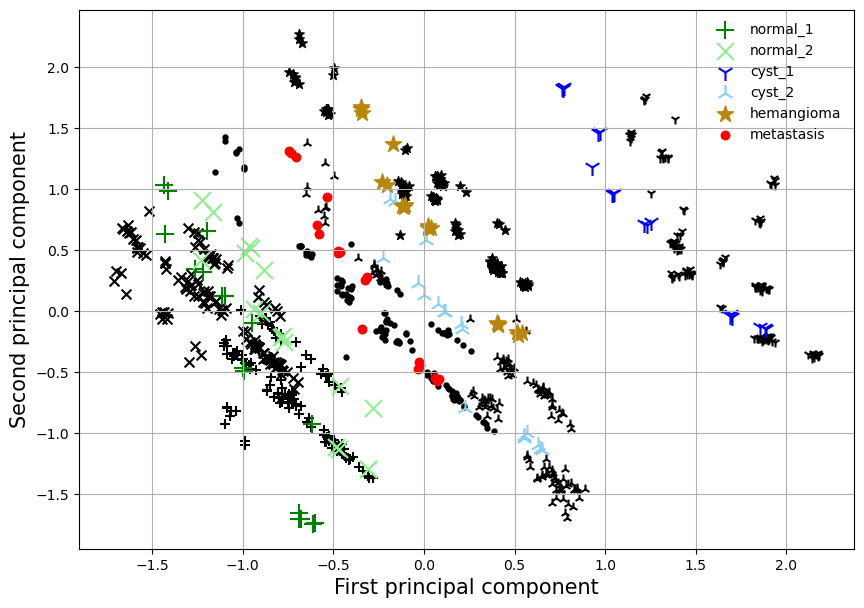}
\caption{Multi-center principal component analysis. The black markers indicate the multi-center empirical phantom study data of the region of interest with the same shape as the colored markers. The colored markers indicate the equivalent result of the simulator with iterative reconstruction.}  
\label{fig:pcamulti}
\end{figure}

Furthermore, the radiomics features of the simulator are compared to the multi-center empirical phantom acquisitions in Figure \ref{fig:pcamulti} with a PCA variability analysis. In the simulator the projection number is fixed to 200 and 250 and the noise range extended to 2.5$\times10^{-3}$ - 2.9$\times10^{-3}$ ($\sigma^2/$pixel in  HU$^2$). This parameter range mimics the fixed slice reconstruction thickness and the extended noise range was used to realise the unknown differences inherent in a multi-center study.
As seen from Figure \ref{fig:pcamulti}, the simulation radiomics variability is in agreement with the empirical results.

\subsection{Discussion}
Experimental comparison showed striking similarity between sensitivity analyses carried out with the anthropomorphic phantom and the CT simulator. Despite the approximations, the CT simulator was able to generate images with very similar characteristics, as quantified by the features studied here, to real images of the phantom.  This is essentially a model whose parameters can be changed to match those observed in phantom studies. This similarity opens different avenues for further investigation and practical opportunities. For example, studies with multi-centre, multi-vendor data sets \cite{learningml} can be effortlessly scaled up and automated.


First, the results suggest that sensitivity analysis for new features or new ways to extract features can be initially performed with a CT simulator. This would substantially reduce the efforts and costs required to study generalization properties of new radiomics features, radiomics analyses and image-based learning techniques to new acquisition settings. This is crucial since this generalization ends up being one of the most notorious obstacles in front of translating new quantitative image analysis technologies to clinical practice. 

While hand-crafted features' stability can in theory still be quantified with phantom studies, this approach remains very limited when it comes to assessing stability of advanced algorithms that extract features in a data-driven way, e.g., neural networks. Phantom studies yield very limited number of images and this inhibits using them for assessing stability of neural network-based feature extraction methods. The simulation study we showed here is a direct solution to this issue. The approach can use any CT image as a "phantom", therefore yields a large number of images to perform accurate stability analysis of such advanced algorithms.

Second, training of learning-based methods can be modified to encourage robustness to variations of imaging characteristics during training. For instance, through extensive data augmentation one can gain robustness to variations in Magnetic Resonance Imaging (MRI) acquisitions of the same contrast \cite{deep2}. As the CT simulator can generate images realistic enough to yield similar sensitivity analyses as an empirical phantom study, one can imagine using such simulators for training of highly robust deep leaning models.

The CT simulator we used here did not consider various details of the acquisition due to simplifications of the system's physical model. Our experiments with more complete models, such as Geant4 \cite{geant4}, showed that using such models is challenging due to the difficulty in replicating a given scan and computation time. Making more accurate simulators more user friendly and faster may improve the quality of the sensitivity analyses.  In addition, a possible extension of this work can be the  application of an automatic segmentation method. Allowing for automated and accurate determination of ROIs, especially useful for the segmentation of liver tissue regions.

\section{Conclusions}
Based on the astratoolbox we have created an environment to reproduce artificial variability on an initial CT-image. The environment was verified to replicate the diversification observed from empirical acquisitions via a principal component analysis, both for intra- and inter-scanner analyses. The methodology and simulational tool can accelerate the creation and testing of stable and discriminative radiomics features. More crucially, this tool can generate realistically variable CT-image datasets for training highly robust deep learning models. 

\bibliography{sample}

\appendix
\section{Best performing radiomics comparison}

\begin{table}[h]
\centering
\begin{tabular}{|l|l|}
\hline
\textit{Stability ($\%$) - simulator} & \textit{Stability ($\%$) - empirical phantom study}   \\
\hline
original glszm SmallAreaLowgreyLevelEmphasis & original gldm LargeDependenceHighgreyLevelEmphasis\\
original gldm LargeDependenceHighgreyLevelEmphasis &original glszm SmallAreaLowgreyLevelEmphasis \\
 original gldm SmallDependenceLowgreyLevelEmphasis & original firstorder Median  \\
 original firstorder Kurtosis &original glcm ClusterShade  \\
 original glcm ClusterShade & original gldm SmallDependenceLowgreyLevelEmphasis  \\
original glcm InverseDifferenceMomentNormalised & original firstorder Mean \\
 original glszm LowgreyLevelZoneEmphasis & original glcm InverseDifferenceMomentNormalised \\
original glrlm ShortRunLowgreyLevelEmphasis & original glrlm ShortRunLowgreyLevelEmphasis \\
original glrlm LongRunHighgreyLevelEmphasis & original glszm LowgreyLevelZoneEmphasis \\
original glcm InverseDifferenceNormalised &original glrlm LowgreyLevelRunEmphasis  \\
\hline
\textit{Discriminative power ($\%$) - simulator} & \textit{Discriminative power ($\%$) - empirical phantom study}\\
\hline

 original firstorder 90Percentile & original firstorder Median \\
 original firstorder Energy & original firstorder Mean \\
  original firstorder Mean & original glszm LargeAreaHighgreyLevelEmphasis \\
  original firstorder Median & original firstorder Energy \\
  original firstorder RootMeanSquared & original firstorder TotalEnergy \\
original firstorder TotalEnergy & original glszm greyLevelNonUniformity \\
  original gldm DependenceNonUniformity &original firstorder Minimum \\
 original glrlm RunEntropy & original glrlm greyLevelNonUniformity \\
 original glszm greyLevelNonUniformity & original glszm LargeAreaLowgreyLevelEmphasis \\
 original glszm LargeAreaEmphasis & original glszm SizeZoneNonUniformity \\
\hline

\end{tabular}
\caption{\label{tab:topfeatures} Comparison table of highest stability and discriminative power radiomics features as observed in the phantom acquisitions and simulational studies. }
\end{table}

\section{Radiomics extraction description \label{app:radiomicsdef}}

The feature definitions are described in the  Imaging Biomarker Standardization Initiative \cite{pyradiomics2}.  More specifically, the radiomics feature categories used are, https://pyradiomics.readthedocs.io/en/latest/index.html, :
\begin{itemize}

  \item  18 first-order statistics which describe the distribution of voxel intensities within a region through common metrics. For example, 
  \begin{equation}
            \textit{energy} = \displaystyle\sum^{N_p}_{i=1}{(\textbf{X}(i)})^2,
  \end{equation}
  which is  a measure of the magnitude (sum of the squares) of voxel values in an image.  $\textbf{X}$ is a set of $N_p$ voxels included in the ROI.

  \item 22 grey level co-occurrence matrices (GLCM). A GLCM of size $N_g \times N_g$ describes the second-order joint probability
  function of an image region constrained by the mask and is defined as $\textbf{P}(i,j|\delta,\theta)$.
  The $(i,j)^{\text{th}}$ element of this matrix represents the number of times the combination of
  levels $i$ and $j$ occur in two pixels in the image, that are separated by a distance of $\delta$
  pixels along angle $\theta$.
  The distance $\delta$ from the center voxel is defined as the distance according to the infinity norm.
  In this work we computed symmetrical GLCM. For example, 
  \begin{equation}
      \textit{autocorrelation} = \displaystyle\sum^{N_g}_{i=1}\displaystyle\sum^{N_g}_{j=1}{p(i,j)ij},
  \end{equation}
  which is a measure of the magnitude of the fineness and coarseness of texture.

  $\textbf{P}(i,j)$ is the cooccurence matrix for an arbitrary $\delta$ and $\theta$.
  $p(i,j)$ is the normalized cooccurence matrix and equal to $\textbf{P}(i,j)/\sum{\textbf{P}(i,j)}$.
  $N_g$ is the number of discrete intensity levels in the image.

  \item 14 Grey level dependence matrices (GLDM). A GLDM  quantifies grey level dependencies in an image.
  A grey level dependency is defined as the number of connected voxels within distance $\delta$ that are
  dependent on the center voxel.
  A neighbouring voxel with grey level $j$ is considered dependent on center voxel with grey level $i$
  if $|i-j|\le\alpha$. In a grey level dependence matrix $\textbf{P}(i,j)$ the $(i,j)$ th
  element describes the number of times a voxel with grey level $i$ with $j$ dependent voxels
  in its neighbourhood appears in image. For example:
  \begin{equation}
      \textit{grey level variance} = \displaystyle\sum^{N_g}_{i=1}\displaystyle\sum^{N_d}_{j=1}{p(i,j)(i - \mu)^2} \text{, where }
      \mu = \displaystyle\sum^{N_g}_{i=1}\displaystyle\sum^{N_d}_{j=1}{ip(i,j)}
  \end{equation}

    Measures the variance in grey level in the image.
   $N_g$ is the number of discrete intensity values in the image.
   $N_d$ is the number of discrete dependency sizes in the image.
   $N_z$ is the number of dependency zones in the image.
  $\textbf{P}(i,j)$ is the dependence matrix.
   $p(i,j)$ is the normalized dependence matrix, defined as $p(i,j) = \textbf{P}(i,j)/N_z$.

\item 16 grey level run length matrices (GLRLM). A GLRLM quantifies grey level runs, which are defined as the length in number of
  pixels, of consecutive pixels that have the same grey level value. In a grey level run length matrix
  $\textbf{P}(i,j|\theta)$, the $(i,j)^{\text{th}}$ element describes the number of runs with grey level
  $i$ and length $j$ occur in the image (ROI) along angle $\theta$. The value of a feature is calculated on the GLRLM for each angle separately, after which the mean of these values is returned.
   For example, grey Level Variance as above. 
    \item 16 grey level size zone matrices (GLSZM). A GLSZM quantifies grey level zones in an image. A grey level zone is defined as a the number
  of connected voxels that share the same grey level intensity. A voxel is considered connected if the distance is 1
  according to the infinity norm (26-connected region in a 3D).
  In a grey level size zone matrix $P(i,j)$ the $(i,j)^{\text{th}}$ element equals the number of zones
  with grey level $i$ and size $j$ appear in image. Contrary to GLCM and GLRLM, the GLSZM is rotation
  independent, with only one matrix calculated for all directions in the ROI. For example, grey Level Variance as above.

\end{itemize} 

\end{document}